\documentclass[12pt,letterpaper]{article}
\usepackage[margin=1in]{geometry} 
\usepackage[singlespacing]{setspace}
\usepackage{amsmath}
\usepackage{amsfonts}
\usepackage{graphicx}
\usepackage{caption}
\usepackage{subcaption}
\usepackage{float}
\usepackage{verbatim}
\usepackage{algorithm}
\usepackage{algorithmic}
\usepackage{multicol}
\usepackage{hyperref}
\renewcommand{\vec}[1]{\mathbf{#1}}
\newcommand{\slfrac}[2]{\left.#1\middle/#2\right.}

\begin{document}

\title{Casimir Torque In Inhomogeneous Dielectric Plates}
\author{William Long and Lucas Schuermann}
\date{}
\maketitle

\begin{abstract}
In this work, we consider a torque caused by the well known quantum mechanical Casimir effect arising from quantized field fluctuations between plates with inhomogeneous, sharply discontinuous, dielectric properties. While the Casimir effect is a relatively well understood phenomenon, systems resulting in lateral or rotational forces are far less developed; to our knowledge, a theoretical study of discontinuous dielectric variants of such systems has not been attempted. We utilize a Proximity Force Approximation in conjunction with the Lifshitz dielectric formula to perform theoretical analyses of resultant torques in systems with bisected and quadrisected dielectric regions. We also develop a high precision Monte Carlo type numerical integrator to approximate our derived expressions. Our calculations of an energy density linear with the alignment angle result in a constant torque and have implications in NEMS (nano electromechanical systems) and MEMS (micro electromechanical systems), including a postulated nanoscale oscillating drive mechanism powered by quantum field interactions.
\end{abstract}

\section{Introduction}
The Casimir Effect, first postulated by Hendrik Casimir over 65 years ago, is one of the most striking manifestations of a quantum electrodynamic (QED) effect on a classical system. Casimir predicted that parallel, conducting plates at micro-scale separations would experience a net attractive force inversely proportional to the fourth power of the separation, a force dominating all others at sub-micron separations [1]. This so-called Casimir force has subsequently been well developed in recent years theoretically and with precise experimental verifications using modern lab equipment such as torsion pendulums and atomic force microscopes (AFM) by prolific researchers like Mohideen and Lamoreaux et al. [2,3]. 
\\\\
Of course, the concept of forces acting on neutral particles is not new in physics: retarded ($\frac{1}{R^7}$) and short range ($\frac{1}{R^6}$) van der Waals potentials can explain attractive forces between neutral atoms. However, the result obtained in weak coupling by deriving the force between plates using these interactions ends in a result only within 20\% of the actual Casimir force. This discrepancy is caused by the fact that van der Waals forces are not pairwise additive due to multiple scattering. The solution to the actual derivation lies in the quantum mechanical phenomena known as zero point energy which, according to quantum field theory, pervades all of space. The concept of zero point energy comes from the realization that the classical view of vacuum: a container emptied of all particles, is not entirely accurate. In reality, vacuums are filled with so-called ``virtual particles'' in a continuous state of fluctuation. Virtual particle-antiparticle pairs are constantly being created and subsequently annihilated with the length of their existence governed by the Heisenberg uncertainty relation:
\begin{align}
\Delta{E}\Delta{t}\geq\hbar
\end{align}
Photons are the dominant particles in a vacuum responsible for creating the measurable Casimir force while other particles provide a subdominant contribution. In general, bosons, including photons, produce a net attractive force while fermions produce a repulsive force, although the nature of the force is dependent on the boundary topology. Furthermore, the nature of the force is also dependent on the surface profile and geometry of the confining plates thus leaving wide room for research expansion in the field.
\\\\
We now provide a brief derivation of the Casimir force from quantum field theory. The Hamiltonian of the electromagnetic field as the sum of independent harmonic oscillators can be written as:
\begin{align}
\hat{H}=\sum_{\kappa}\sum_{\lambda}{\hbar\omega_\kappa(\hat{a}_{\kappa\lambda}^{\dagger}}\hat{a}_{\kappa\lambda}+\frac{1}{2})
\end{align}
By defining the vacuum state as the state with no photons of any mode, the vacuum energy can be written as:
\begin{align}
E_0=\frac{1}{2}\sum_{\kappa}\sum_{\lambda}\hbar\omega_{\kappa}=\sum_{\kappa}\hbar\omega_{\kappa}
\end{align}
￼We now introduce two parallel perfectly conducting plates into the system, thereby creating a cavity in the vacuum and establishing Dirichlet and Neumann boundary conditions on the wave modes between the plates. Thus, only electromagnetic waves with modes on both walls of the plates can exist ($\varphi\rvert_{\partial V}=0$). Discrete modes can exist within the cavity with wave numbers:
\begin{align}
k=\frac{\nu\pi}{L}
\end{align}
where $L$ is the separation distance and ${\nu \in \mathbb{N}, \nu \neq {0}}$.
Thus, by knowing that in a vacuum, the angular frequency is related to the wave number by the speed of light, we can rewrite the zero point energy with the relation ${\omega=ck}$ as:
\begin{align}
E_0=\frac{\pi\hbar{c}}{L}\sum_{\nu=1}^\infty{\nu}
\end{align}
￼Here, we find a curiosity in that although the energy, which is obviously divergent, goes to infinity since we are summing over all possible electromagnetic modes; a finite meaningful result can be obtained from subtracting the zero point energy within the cavity from the outer vacuum energy. This difference represents the imbalance between radiative pressure from continuous modes outside the cavity and discrete modes within the cavity. We then obtain the following:
\begin{align}
\Delta{E}=E_0-E_c=\frac{\pi\hbar{c}}{L}[\sum_{\nu=1}^\infty{\nu}-\int_0^\infty\nu{d}\nu]
\end{align}
The sum and the integral here are divergent; therefore, we introduce a smooth cutoff function $\Lambda (\slfrac{k}{k_c})$ having the properties: $\Lambda(\slfrac{k}{k_c})\to1$, $k \ll k_c$ and $\Lambda(\slfrac{k}{k_c})\to0$, $k \gg k_c$. Now, implementing the Euler-MacLaurin summation formula and a conversion factor $\lim_{\epsilon \to \infty}{e^{-\epsilon\nu}}$ we have the energy: 
\begin{align}
E_{cas}=-\frac{\pi^2}{720}\frac{\hbar{c}}{L^3}A
\end{align}
￼
Along with the famous net attractive Casimir force:
\begin{align}
F_{cas}=-\frac{\pi^2}{240}\frac{\hbar{c}}{L^4}A  
\end{align}
where ￼$\hbar$ is Planck’s constant, ￼$c$ is the speed of light in vacuum and￼ $A$ is the area of the plates. Or else, in systems arising in rotational forces, we obtain:
\begin{align}
\tau_{cas}=-\frac{\partial E_{cas}}{\partial \alpha}
\end{align}
where $\alpha$ is the alignment angle between the bifurcated plates. The plates attract or rotate with respect to each other in order to minimize the energy associated with the field fluctuations and, in doing so, generate a measurable restoring force or torque. Of course this brief derivation only considers the one-dimensional case, but the result can be easily extended to three or more dimensions. While we utilize the zero point energy derivation, the method Hendrik Casimir originally used to obtain his answer, most modern derivations make use of more advanced mathematical transformations such as zeta function and exponential regularization [4, 5].
\\\\
The Casimir force has implications reaching far beyond just field theory. Because of its topological dependence, the force can imply a closed or open universe and the number of space-time dimensions in cosmology [6]. The Critical Casimir effect, the thermodynamic analogue of the quantum mechanical effect, plays an important role in the surface and bulk critical phenomena of condensed matter physics [7]. Also in mathematical physics, the Casimir effect has prompted the development of regularization and renormalization techniques based on zeta function and heat kernel expansion [8]. In NEMS and MEMS, stiction --- the collapse of movable elements into nearby surfaces as a result of Casimir interactions --- has led many to believe that there may be a fundamental limit on how small nano-devices can be made [9]. On the other hand, a vast array of applications have arisen from these interactions including a theoretical ``rack and pinion'' device and postulated nano-scale drive mechanisms [10, 11].
\\\\
\begin{figure}[H]
\centering
\includegraphics[width=.5\textwidth]{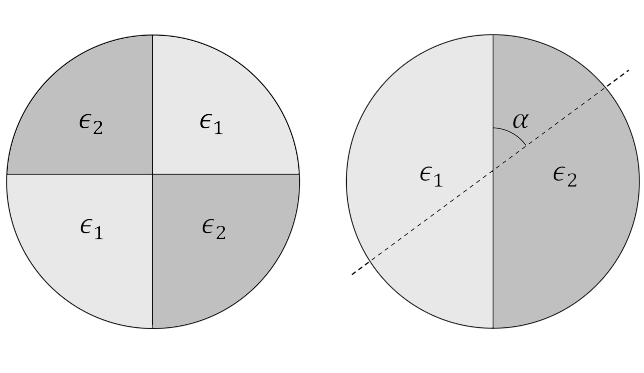}
\caption{Diagram of dielectric plate system configurations (System A: right, System B: left) with quadrisected and bisected dielectric regions where the dashed line represents the dividing line between the dielectric regions for the top bifurcated disk (not shown), and $\alpha$ is the alignment angle.}
\end{figure}
\begin{figure}[H]
\centering
\includegraphics[width=.4\textwidth]{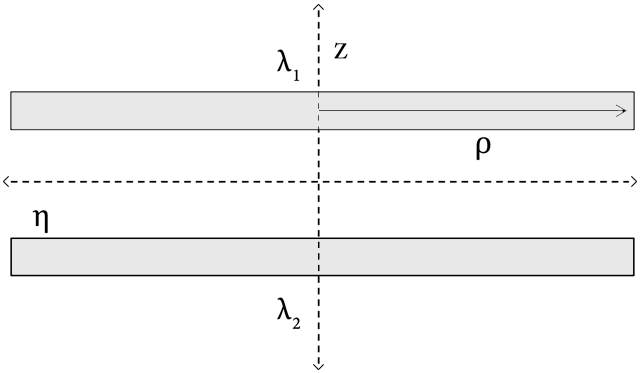}
\caption{Diagram from a side view of our dielectric plate system where $\lambda_1, \lambda_2$ are the coupling constants of the top and bottom plates, respectively, $\rho$ is the radius of the disks, $z$ is the radial axis, and $\eta$ is the lateral axis.}
\end{figure}
Due to our motivation for a practical application, we are forced to depart from the majority of work in this field. While the vast body of work in theoretical Casimir physics pertains to perfectly conducting plates for ease of calculation, our present work addresses plates of finite conductivity with realistic dielectric properties. Lifshitz, Dzyaloshinskii and Pitaevskii were the first to calculate the Casimir interactions for plane dielectric surfaces [12] while Bordag et al. presented a definitive work on the Casimir effect in finite dielectric systems [13]. These systems are more difficult to analyze and require a slightly different formulation to account for finite conductivity. Also, less work has been done on rotational and lateral forces arising from the Casimir effect as opposed to perpendicular forces. Therefore, this category of Casimir configurations is a far more recent field of study relative to general Casimir physics with much less research and development overall. In this work, we consider two unique sectioned dielectric plate systems. The first and simpler configuration for these systems, System $A$ (see Fig. 1), is composed of two bifurcated finite disks each bisected into two even areas, each area with a different dielectric property. The two disks are initially positioned at a short distance $L$ and aligned in such a way that the dividing lines of the dielectric regions of one plate create an angle $\theta=\frac{\pi}{4}$ with the lines of the mirroring plate. Our second configuration, System $B$, is identical to $A$ but with the disks quadrisected into four regions, each region pair opposite the axis having the same dielectric property for a net total of two different dielectric regions.
\\\\
Of the existing body of work in rotational forces, the majority of studied systems involve structured corrugations within the plates. These generally periodic corrugations have included saw tooth and sine wave profiles among others, resulting in a calculable torque [14, 15, 16]. The configurations we consider provide a distinct advantage over other systems in that at the micron scale, at which these plates are constructed in NEMS and MEMS, producing nano-scale corrugations on a plate cannot be accomplished through either wet or dry chemical etching. Manufacturing these fine profiles at high accuracy and keeping corrugated plates clean of obstructing particles, the main reason why Casimir effect experiments generally take place in vacuum, make these systems difficult to create and practically infeasible to utilize. On the other hand, our systems do not require any particular fine structuring. Instead, integrating two half planes of different materials can be easily accomplished using modern instrumentation and provides, arguably, the only viable means to synthesize a system that results in a nontrivial torque. Another field of recent work in rotational and lateral forces has focused on birefringent plates, plates with in-plane optical anisotropy [17, 18]. A spontaneous restoring torque can arise in these plates towards the configuration in which their principal axes are aligned. However, there are relatively few optically anisotropic materials, most of which are incompatible with nano-device construction, whereas our dielectric systems allow for any arbitrary dielectric material. Therefore, our systems of bisected and quadrisected dielectric plates seem to be the most suitable option for incorporation into various NEMS applications.
\\\\
While systems of varying dielectric properties have been investigated, they have been limited to brief works on plates with piecewise continuous dielectric profiles. Our systems of sharply discontinuous profiles resulting in Casimir torques have, to our knowledge, never been formally investigated by researchers in this field. We endeavor, in studying this particular system, to calculate the energy field density and use it to derive a numerical torque dependent on the choice of dielectric properties, separation distance, plate length, temperature, and surface profile. Therefore, this work represents the first attempt at understanding a Casimir system fundamental to the development of increasingly advanced nanotechnologies.

\section{Methods}
\subsection{Formulations}
To study our dielectric plates, we utilize Lifshitz's approach with the Proximity Force Approximation (PFA), an analytic formulation making several assumptions regarding the system. In Lifshitz theory, all magnetic properties are neglected with the magnetic permeability set equal to 1 for all frequencies [19]. Furthermore, the PFA assumes that any small differential area $dxdy$ of one plate interacts only with the corresponding area directly across from it on the opposite plate. Of course, in reality, subdominant interactions occur between each area with every other area on both plates, but the approximation assumes these corrections to be negligible. This assumption allows for a simplification of the solution technique.
\\\\
In general, the interaction energy of the system in weak coupling can be written using the direct formulation:
\begin{align}
E&=-\frac{1}{64\pi^3}\int{d\vec{r}^3d\vec{r'}^{3}\frac{V_1(\vec{r})V_2(\vec{r'})}{{|\vec{r}-\vec{r'}|}^3}} \\
\nonumber  V_1&=4\pi\, \delta(z-\lambda_1)[\theta(\varphi+\frac{\pi}{2})\theta(\frac{\pi}{2}-\varphi)\lambda_1+\theta(\varphi-\frac{\pi}{2})\theta(\frac{3\pi}{2}-\varphi)\lambda_2]\theta(-\rho+\eta) \\
\nonumber  V_2&=4\pi\, \delta(z-\lambda_2)[\theta(\widetilde{\varphi}+\frac{\pi}{2})\theta(\frac{\pi}{2}-\widetilde{\varphi})\lambda_1+\theta(\widetilde{\varphi}-\frac{\pi}{2})\theta(\frac{3\pi}{2}-\widetilde{\varphi})\lambda_2]\theta(-\rho+\eta)
\end{align}
where $V_1$ and $V_2$ are the interaction potentials of the respective plates, $\lambda_1, \lambda_2$ are the coupling constants, $\varphi$ is the angle with $\widetilde{\varphi} = \varphi + \alpha$, $\delta$ is the Dirac delta function, and $\theta$ is the Heaviside step function. This formula is for a scalar field, so $\lambda$ has dimensions $\frac{1}{L}$. Instead of attempting a brute force analytic approach which ends up involving massively complex multiple integrals, we can solve the Lifshitz double integral for the energy density in each vacuum region between $\epsilon_1$ on the first plate and $\epsilon_3$ on the opposite plate with some dielectric $\epsilon_2$ in between them using the formula:
\begin{align}
\mathcal{L}=\int_{0}^\infty{d\zeta}\int_{\zeta}^\infty{d\kappa\cdot\kappa[\ln(1-\Delta^{TM})+\ln(1-\Delta^{TE})]} \\
\nonumber \Delta^{TE}=1-r_1^Er_2^Ee^{-2\kappa{a}} \quad \Delta^{TM}=1-r_1^mr_2^me^{-2\kappa{a}}
\end{align}
where $a$ is the separation distance of the plates, $r_1$, $r_2$ are the reflection coefficients of their respective plates derived from multiple scattering and $\Delta^{TE,TM}$ are the transverse electric and transverse magnetic contributions to the energy respectively. In this equation, the dielectric properties are assumed to be functions of momentum and imaginary frequency according to the Drude Model [20]; that is: $\epsilon_a=\epsilon_a(i\zeta,\rho)$ where $\zeta$ is the frequency. The reflection coefficients are defined to be:
\begin{align}
r_1^Er_2^E&=(\frac{\slfrac{\kappa_1}{\epsilon_1}-\slfrac{\kappa_2}{\epsilon_2}}{\slfrac{\kappa_1}{\epsilon_1}-\slfrac{\kappa_2}{\epsilon_2}})(\frac{\slfrac{\kappa_3}{\epsilon_3}-\slfrac{\kappa_2}{\epsilon_2}}{\slfrac{\kappa_3}{\epsilon_3}-\slfrac{\kappa_2}{\epsilon_2}}) \\
r_1^mr_2^m&=(\frac{\kappa_1-\kappa_2}{\kappa_1+\kappa_2})(\frac{\kappa_3-\kappa_2}{\kappa_3+\kappa_2})
\end{align}
Here, $\kappa$ is the wave number $\kappa_a=\sqrt{\kappa_\perp^2+\zeta^2\epsilon_a(i\zeta,\rho)}$ where $\kappa_\perp$ is the propagation vector in the transverse direction. From here, we optimize the formulation by first assuming that the dielectric properties are constant or independent of frequency and momentum for our purposes. Next, we take the variable transformations $x=2\kappa{a}$ and $y=2\zeta{a}$. Thus, the following formulation is obtained:
\begin{align}
E&=\frac{1}{32\pi^2a^3}\int_{0}^\infty{dy}\int_{y}^\infty{dx} {\cdot} \\
\nonumber&\qquad x\left( \ln{}[1-e^{-x}(\frac{x-\sqrt{x^2+(\epsilon_1-1)y^2}}{x+\sqrt{x^2+(\epsilon_1-1)y^2}})\cdot(\frac{x-\sqrt{x^2+(\epsilon_2-1)y^2}}{x+\sqrt{x^2+(\epsilon_2-1)y^2}})]\right. \\
\nonumber&\qquad +\left. \ln{}[1-e^{-x}(\frac{\epsilon_1x-\sqrt{x^2+(\epsilon_1-1)y^2}}{\epsilon_1x+\sqrt{x^2+(\epsilon_1-1)y^2}})\cdot(\frac{\epsilon_1x-\sqrt{x^2+(\epsilon_2-1)y^2}}{\epsilon_1x+\sqrt{x^2+(\epsilon_2-1)y^2}})]\right)
\end{align}
To obtain the final energy configuration for our plate system, we obtain the following function of the alignment angle $\alpha$ from the PFA of Fig. 2
\begin{align}
E_{PFA,\,2}&=\mathcal{L}^{1,1}\frac{\pi-\alpha}{2}\rho^2+2\mathcal{L}^{1,2}\frac{\alpha}{2}\rho^2+\mathcal{L}^{2,2}\frac{\pi-\alpha}{2}\rho^2\\
\nonumber E_{PFA,\,4}&=2\mathcal{L}^{1,1}\frac{\alpha}{2}\rho^2+4\mathcal{L}^{1,2}\frac{\slfrac{\pi}{2}-\alpha}{2}\rho^2+2\mathcal{L}^{2,2}\frac{\alpha}{2}\rho^2
\end{align}
\subsection{Numerical Analysis Techniques}
To verify the results of our analytical derivation and subsequent numerical analysis using Wolfram Mathematica [21], a high precision numerical integrator is needed. When working on a simulation of the Casimir effect between arbitrary geometries using the Worldline approach, Gies et. al. successfully implemented a numerical integration method based on Monte Carlo techniques [22]; due to the nature of the equations, standard numerical integration techniques are limited in their applications to our formulations, thus we also choose to implement a Monte Carlo type integrator.
\\\\
We will now provide a brief derivation and discussion of our use of the Monte Carlo techniques. Monte Carlo integration is a numerical technique to estimate a definite integral using random sampling - as opposed to the grid-based sampling often used by other techniques - that is specifically well suited to evaluating multidimensional integrals. In accordance with Weinzierl's discussion of Monte Carlo methods in high energy physics [23], the Monte Carlo estimate for the integral 
\begin{align}
I=\int{f(x)dx}
\end{align}
where $f(x)=f(u_1,...u_d)$ is given by
\begin{align}
E=\frac{1}{N}\sum_{n=1}^{N}f(x_n).
\end{align}
The law of large numbers ensures that the Monte Carlo estimate converges to the true value  of the integral, that is, 
\begin{align}
\lim_{N \to \infty}\frac{1}{N}\sum_{n=1}^{N}f(x_n) = I.
\end{align}
Monte Carlo methods are inherently parallel: the performance of a computer implementation of this integration technique can be drastically increased by the use of multiprocessing. For example, Eq. 17 is particularly well suited to a parallel implementation, as the calculation of $f(x_n)$ for a particular $x_{n_0}$ can be completed concurrently to the calculation of $f(x_n)$ for another given $x_{n_1}$. Due to this observation, our integrator was written in Python to work on multiple CPU cores in parallel, leveraging the parallel nature of the algorithm to allow for faster processing times. The estimation of the error of $E$ is given by
\begin{align}
\delta E \approx \sqrt{\text{Var}(E)} = \frac{\sigma_N}{N}
\end{align}
where $\sigma_N$ is the standard deviation of the sampling points, $x_n$. It is apparent that the error bound decreases in relation to the value of $\frac{1}{\sqrt{N}}$ as $N$ increases, thus our implementation seeks to maximize the number of sampling points while maintaining feasible computing time. A straightforward algorithm for Monte Carlo integration of an arbitrary function can be written as follows:
\begin{algorithm}[h!]
\caption{Evaluates $f(x)$ using Monte Carlo integration}
\begin{algorithmic}[1]
\REQUIRE {Define an arbitrary $integrand$ of $f(x)$, the number of iterations, $N$, and the lower and upper bounds for $x$, $xmin$ and $xmax$}
\REQUIRE Let $sample(n)$ return a uniform random sample $n$ between $xmin$ and $xmax$ such that the range of the random samples covers the domain of $x$. This functionality is provided in many modern programming languages, such as Python's $random.uniform$ and $itertools.islice$.
\STATE $domain = xmax - xmin$
\STATE $sum = 0$
\STATE $sum\_squared = 0$
\FOR{all $n$ from 1 to $N$}
	\STATE $x = sample(n)$
	\STATE $f = integrand(x)$
	\STATE $sum = sum + f$
	\STATE $sum\_squared = sum\_squared + f^2$
\ENDFOR
\STATE $mean = \slfrac{sum}{N}$
\STATE $variance = (sum\_squared-\slfrac{(\slfrac{(\slfrac{sum}{N})^2}{N})}{(N-1)})$
\STATE $estimate = domain\cdot mean$
\STATE $error = domain\cdot \sqrt{\slfrac{variance}{N}}$
\RETURN $estimate$, $error$
\end{algorithmic}
\end{algorithm}
\\
Note that this algorithm can easily be parallelized by executing each iteration of the loop independently on a different processing core. A direct Python implementation of this algorithm resulted in an average loop execution time of approximately $10 \mu{s}$, meaning that with a time constraint of $12$ hours, over $4$ billion points could be sampled even without threading optimization. After writing the code to leverage all 8 cores of the host computer, we observed a performance increase of approximately 10x, from computation times of 47.79 seconds with the serial version to 4.60 seconds with the multiprocessor revision (1 million iterations were used in these test cases).

\begin{figure}[H]
\centering
\includegraphics[width=0.9\textwidth]{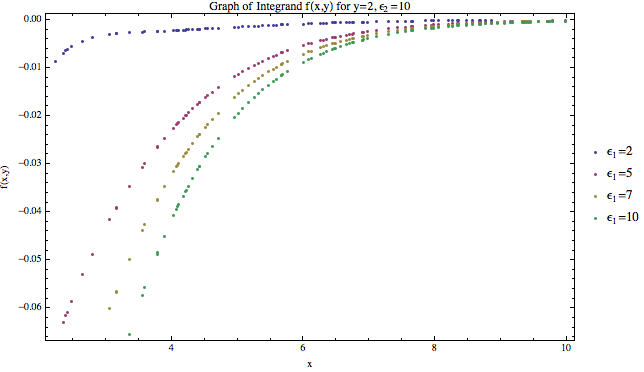}
\caption{Monte Carlo sample points of the integrand for varying dielectric constants. The value of $y$ is fixed for ease of visualization of the method.}
\end{figure}
\section{Results and Discussion}
\begin{figure}[t]
\centering
\includegraphics[width=0.9\textwidth]{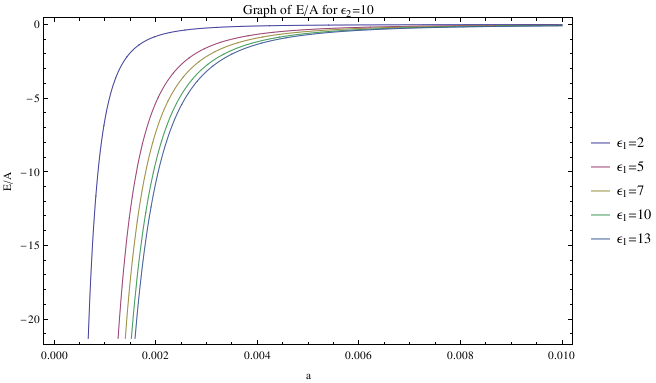}
\caption{Energy densities of the two plate system as the plate separation varies for $\epsilon_2=10$ and different values of $\epsilon_1$ (see key).}
\end{figure}
Beginning with our formulation in Eq. (14), we define the function $f(y)$ to be:
\begin{align}
f(y)&=\int_{y}^\infty{dx} {\cdot}x[\ln({1-\Delta^{TE}})+\ln({1-\Delta^{TM}})]
\end{align}
We then utilize the Monte Carlo numerical integrator to accurately obtain values of $f(y)$ for arbitrary choices of the lower bound $y$. An illustration of the Monte Carlo sampling of the integrand for this calculation is shown in Fig. 3. These results are displayed with an array of arbitrary $\epsilon$ to convey asymptotic behavior of the dielectric property dependence [24]. It is evident from the graph that the $x$ integration converges very quickly to zero.
Following the evaluation of the $x$ integral, we then move on to the outer $y$ integral requiring extensive time and processing power. We define the function $g(a)$ to be:
\begin{align}
g(a) = \frac{1}{32\pi^2a^3}\int_{0}^\infty{f(y)}dy
\end{align}
Values for this integral are found by computing various values of $f(y)$ for sample choices of $y$. With 250  million iterations of the Monte Carlo computation, we were able to calculate a value of $g(a)$ to within $0.33755\%$ of Mathematica's result with an estimated error bound of $0.01475$. Multiplying our result by a factor of $\frac{\hbar c}{32\pi^2}$, we finally gain the function for the final energy density of the system for an angle $\alpha$ with a $\frac{1}{a^3}$ dependence on the separation distance as seen in Fig. 2.
\\\\
\begin{figure}[t]
\centering
\includegraphics[width=0.9\textwidth]{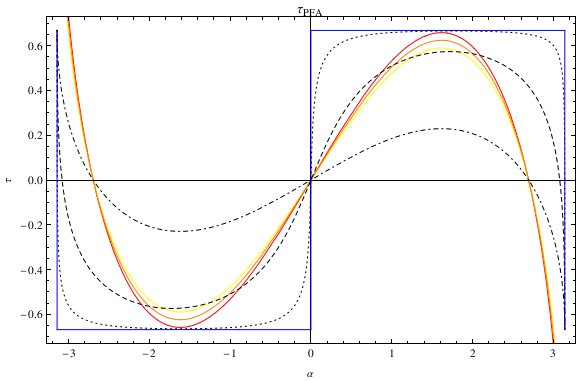}
\caption{Our calculated PFA estimate of Casimir torque for System A (shown in blue) plotted against possible representations of the analytically calculated torque as a function of the alignment angle $\alpha$. We utilize B\'ezier curves to obtain these representations with the points of discontinuity in the PFA torque as argument values. The colored graphs display curves with vertically shifted arguments while  the dotted graphs show the initial Bezier curve, and curves with 8 and 16 times reiterated arguments exhibiting increasing convergence to the PFA torque. The curve with 16 reiterations is a relatively accurate representation of the analytical torque. This theoretical curve would be completely piecewise continuous on $[-\pi,\pi]$ with smooth transitions between -$\tau$, 0, and $\tau$.}
\end{figure}
Finally, choosing an arbitrary $a=0.01, R=1$ where R is the radius of the two bifurcated disks results in an array of values of $\mathcal{L}^{a,b}$ for our choices of $a$ and $b$ that we then use to compute Eq. 15. $E_{PFA}$ is the final result for the energy configuration according to the Proximity Force Approximation and has a simple linear dependence on the alignment angle $\alpha$. Using this formulation, the value obtained is valid only for the interval $[-\pi,\pi]$ but is also $2\pi$ periodic allowing for extensions to any angle $\alpha$. An exact analytical solution would provide a small correction to this simple behavior. The resultant torque for System A, $\tau_{PFA, \, 2}$ is obtained using Eq. 9 and is displayed in Fig. 5. Our results for Systems A and B with arbitrary choices $\epsilon_1 = 8,\, \epsilon_2 = 10,\, r = 1,\, a = 0.002$ evaluated to $0.66819$ and $1.3364$ respectively. These results infer that the Casimir torque increases with an increasing number of sectioned dielectric regions. We then utilize B\'ezier curves to provide insight into how an analytically solved solution would appear. B\'ezier curves with increasing point iterations approach the graph of the solution and exhibit the smooth behavior of the analytic solution. These curves show that it is necessary to transcend the PFA approximation with weak coupling or an exact analytic solution.

\section{Conclusion and Future Work}
In conclusion, we have presented a theoretical study of a Casimir torque arising from a system of bisected and quadrisected inhomogeneous dielectric plates. By means of a high precision Monte Carlo type integrator, we were able to utilize the Lifschitz dielectric formula with several simplifying assumptions to obtain the energy density of the system. Finally, the Proximity Force Approximation allowed us to derive the resultant torque. Through this research, we contribute to the growing body of work on the rotational and lateral Casimir Effect while offering a unique application in the field of nanotechnology. 
\\\\
These applications include devices such as the nano-scale drive mechanism discussed previously. If such a drive mechanism were used to drive a nano-device, the magnitude of the Casimir interactions could easily be tuned to the requisite specifications using the parameters we consider in this work. A large driving torque along with relatively low damping from small scale machine gears would result in a near perpetually running rotor. The prospect of powering a nano-device using a renewable quantum mechanical effect is certainly exciting although more engineering analysis would be required. 
\\\\
While recent work has focused on corrugated and birefringent plates, the dielectric plate system we introduce provides a distinct advantage in its relative ease of production and ultra low dampening.  
Due to time constraints, we were forced to make several simplifying assumptions that in future work, could be omitted and compared to our current results. We assumed a frequency independence in $\epsilon$ that, while not of great consequence to our results, could be incorporated in future work on the subject. Additionally, our formulations were all done at zero temperature $T=0K$ and assuming a perfectly smooth plate. In thermal equilibrium, Green's functions are periodic in imaginary time with period $\beta \sim \frac{1}{kT}$ that would allow the calculation of a finite temperature correction. To account for a rough, imperfect surface profile, a profile generator utilizing a random number generator could be used to provide extremely small, but more accurate, corrections. These corrections, if calculated, could be easily appended to the results of this work to provide insight into more realistic conditions. Furthermore, because of the infeasibility of solving for the torque exactly, we used a numerical approach in calculating our results. An analytic solution, while nearly impossible with current methods, would likely provide interesting insight into the system and a more generalized result. It would also produce slight corrections to the simple linear behavior of the energy and thus, also the constant torque. An experimental verification would be a more sophisticated continuation of the present work requiring access to expensive, advanced equipment but if pursued would pave the way for incorporation of our research into a multitude of physical applications. 

\section*{Acknowledgements}
We would like to thank Dr. Kimball Milton for instruction and helpful discussions along with the Homer L. Dodge Department of Physics and Astronomy at the University of Oklahoma for providing research facilities, and lastly the Julian Schwinger Foundation and the US National Science Foundation under PHY 1062774 for helping fund this research.
\\\\
*Electronic mail addresses:\\
Lucas Schuermann - \url{lschuermann@ou.edu}\\
William Long - \url{wlong@ou.edu}
\pagebreak


\begin{thebibliography}{1}
\bibitem{24}It should be noted that for simplicity we choose an arbitrary value of $\epsilon_2$ to be 10 for all of our calculations and results
\end{thebibliography}
\end{document}